\documentclass[journal,transmag]{IEEEtran}

\usepackage{cite}
\usepackage{graphicx}
\usepackage{amssymb,amsmath}

\begin{document}

\title{Neural Network Equalization for\\ Asynchronous Multitrack Detection in TDMR}


\author{
	\IEEEauthorblockN{Elnaz Banan Sadeghian\IEEEauthorrefmark{1},~\IEEEmembership{Member,~IEEE}}
	\IEEEauthorblockA{\IEEEauthorrefmark{1}Department of Electrical and Computer Engineering, Stevens Institute of Technology \\Hoboken, NJ 07030, USA, ebsadegh@stevens.edu}
}

\IEEEtitleabstractindextext{%
\begin{abstract}
The advent of multiple readers in magnetic recording opens the possibility of replacing the current industry's single-track detection with the more promising multitrack detection architectures. We have proposed a first solution, a generalized partial-response maximum-likelihood (GPRML) architecture, that extends the conventional PRML paradigm to jointly detect multiple asynchronous tracks. In this paper, we propose to replace the conventional communication-theoretic multiple-input multiple-output equalizer in the GPRML architecture with a neural network equalizer for better adaption to the nonlinearity of the underlying channel. We evaluate the proposed equalization strategy on a realistic two-dimensional magnetic-recording channel, and find that the proposed equalizer outperforms the conventional linear equalizer, by a $35\%$ reduction in the bit-error rate. 
\end{abstract}

}

\maketitle

\pagestyle{empty}
\thispagestyle{empty}

\IEEEpeerreviewmaketitle

\section{Introduction}

\IEEEPARstart{M}{ultitrack Detection} significantly improves the areal density as well as the throughput over the conventional single-track detection schemes. The architecture of a multitrack read channel, however, is substantially different from the single-track read channel, specifically, when it comes to the synchronization component. The reason is that unlike the single-track detection, we can not synchronize any readback waveform to the rates of multiple tracks that are themselves asynchronous. Consequently, we can no longer synchronize the waveform(s) before we can detect them. Instead, the functions of synchronization and detection should be fused together \cite{EBS_thesis}. To this end, we have proposed a generalized partial-response maximum-likelihood (GPRML) architecture for joint detection of multiple asynchronous tracks \cite{APRTV_ICC, GPRML_TCOM}. The proposed architecture consists of an asynchronous partial-response (APR) equalizer to time-varying target, followed by the joint detector of multiple asynchronous tracks, namely the ROTAR detector of \cite{ROTAR_JSAC}. 

On the other hand, the two-dimensional magnetic recording (TDMR) channel exhibits nonlinearity in the jitter-like transitioning and widening of the reader response which is modeled by the dominant and pattern-dependent media noise, in the partial erasure, and also in the two-dimensional interference, to name a few. Hence, neural networks have been proposed since 1990's for equalization and detection in TDMR \cite{Nair_nn1, Nair_nn3, Ober_nn}. Recently, however, neural networks have shown a huge success in replacing the communication-theoretic modules for TDMR read channels \cite{Shen_CISS, Sayyafan_nn, Siegel_RNN}. Nevertheless, no prior work has developed neural networks for multitrack detection, much less for multitrack detection of asynchronous tracks.

In this paper, we propose a neural network multiple-input multiple-output (MIMO) equalizer to replace the linear equalizer in our GPRML read channel. Compared to the linear equalizers used in PRML read channels, the proposed equalizer better adapts the nonlinear channel response and thereby helps mitigate the media noise. Further, the proposed neural network equalizes the unsynchronized readback samples to a time-varying target that absorbs the timing asynchrony between the tracks of interest.

\vspace{-1em}
\section{Neural Network Asynchronous Partial-Response Equalizer to Time-Varying Target}

\begin{figure}[!t]
  \centering
  \includegraphics[width=0.5\textwidth]{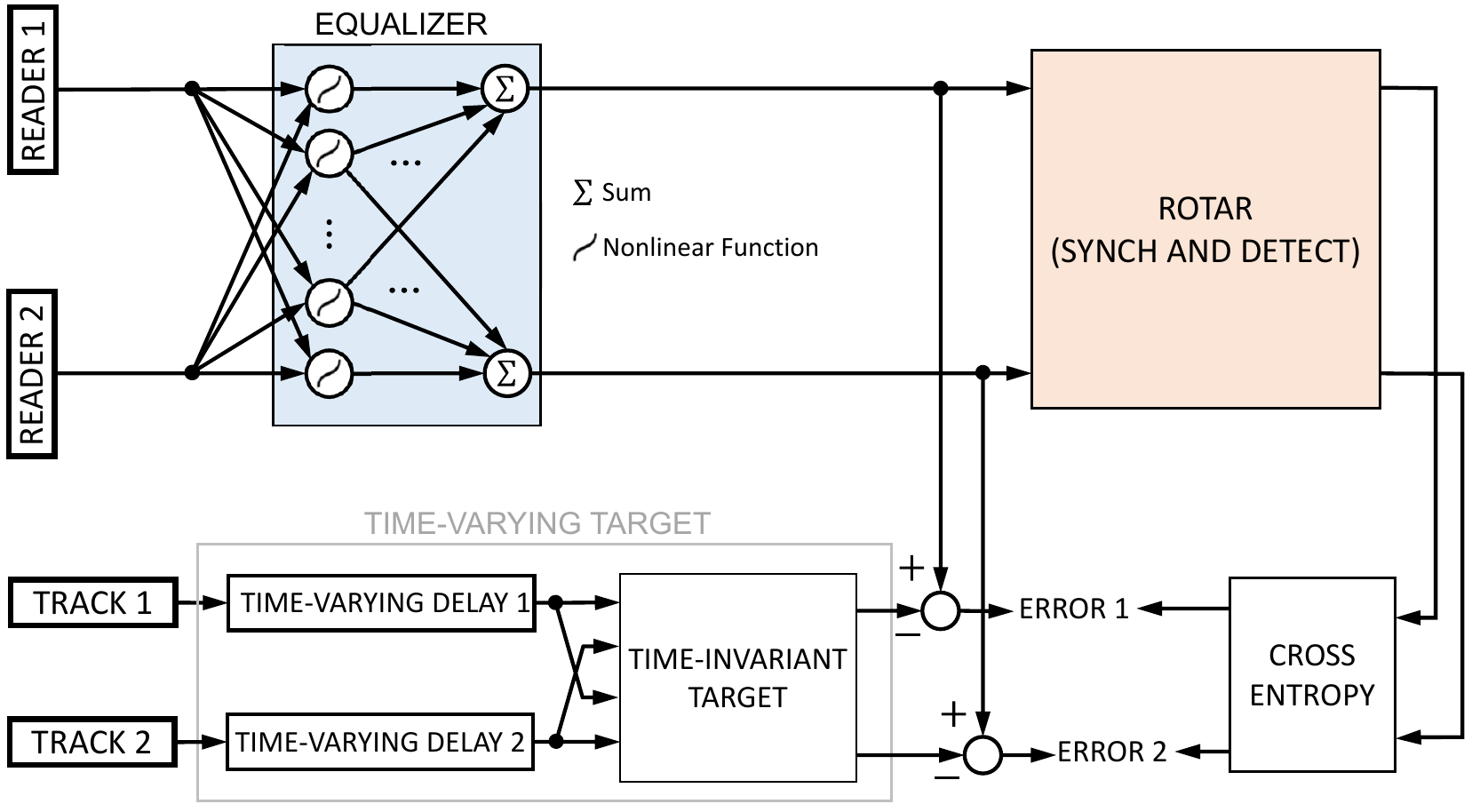}
  \caption{Our proposed MIMO neural network equalization to a time-varying target for the exemplary case of detecting $2$ tracks from $2$ readback waveforms.}
  \label{fig:multitrack-equalization}
  \vspace{-1pt}
\end{figure}

Fig. \ref{fig:multitrack-equalization} illustrates our proposed equalization strategy for an exemplary case of jointly detecting two tracks from two readback waveforms. The upper branch has the MIMO equalizer with one input layer, three fully connected hidden layers (only one layer is shown), and a linear output layer. The input layer feeds the readback samples corresponding to the two readers to all the neurons in the first hidden layer. The hidden neurons apply the tangent-sigmoid activation function to their inputs, and the two neurons of the output layer linearly combine their inputs to form the two equalized outputs corresponding to the two readback waveforms in this figure. The equalized waveforms are then fed to the ROTAR detector for the final detection of the written bits. 

The lower branch shows the strategy for computing the unknown equalizer and the target pair, and the unknown timings, during the training where the user bits are known. The resulted equalizer and target pair are used during the testing with the unseen bits.
The lower branch is adapted from our prior results in \cite{APRTV_ICC, GPRML_TCOM}. According to our prior findings for sufficiently slow-varying timing offsets, the asynchrony in ADC outputs can be accurately modeled by applying a time-varying fractional delay to the user bits when computing the optimum equalizer and the target pair. Therefore, we compute our equalizer and the target pair to work with the unsynchronized ADC samples by replacing the original bits with fractionally delayed bits. Accordingly, in Fig. \ref{fig:multitrack-equalization}, the bits on tracks $1$ and $2$ are first passed through two interpolation filters that apply fractional delays equal to the estimated timing offsets of the two tracks. Next, the fractionally delayed bits are modulated with a matrix-valued target response. The cascade of the fractional delay filters with the time-invariant target responses can be viewed as a time-varying target. 

We use the backpropagation algorithm to optimize the neural network connection weights and the target coefficients along side a second-order phase-locked loop to compute the fractional delay elements. We have used the minimum squared error between the upper and the lower branch, as shown in Fig. \ref{fig:multitrack-equalization}, as well as the cross-entropy between the actual bits and the ROTAR detector's estimates of the bits as our loss functions. The cross-entropy leads to better bit-error rate (BER) results and therefore it is implemented for the simulations results, as follows.
\vspace{-1em}
\section{Simulation Results}

\begin{figure}[!t]
\centering
\includegraphics[width=3.3in]{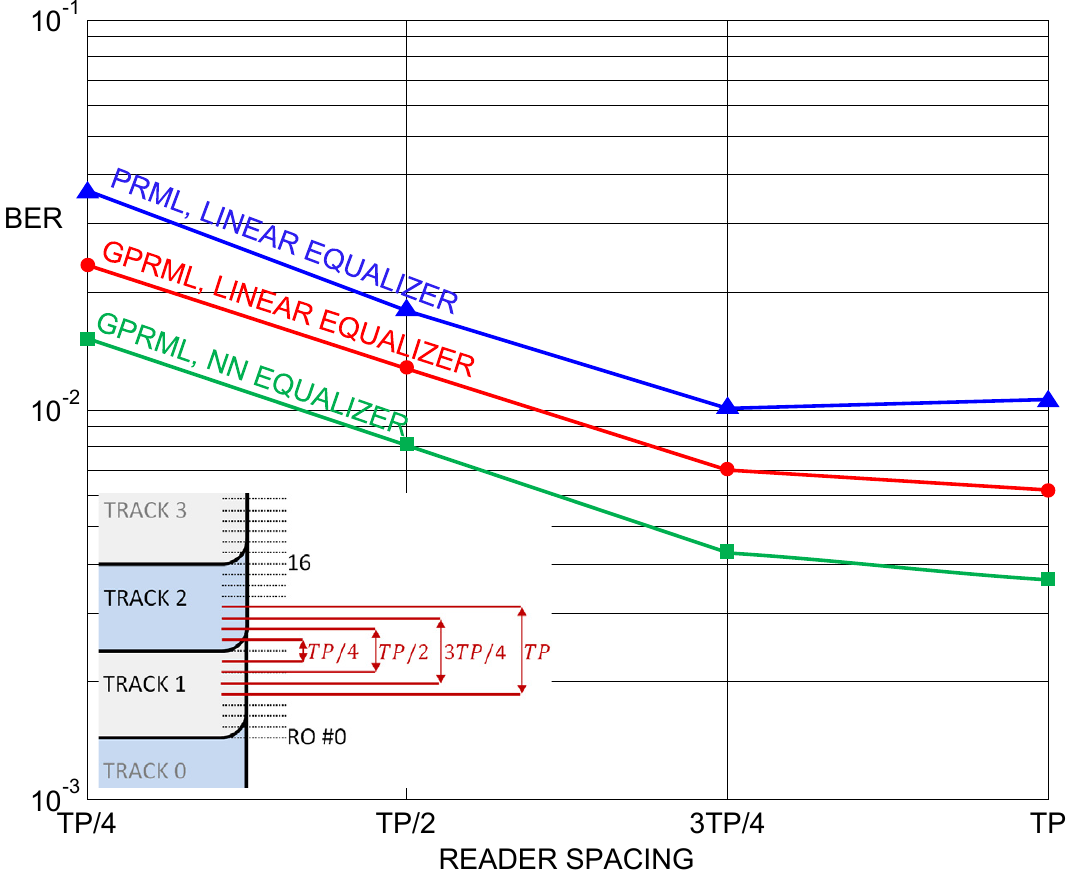}
\vspace*{-0.5em}
\caption{BER performance of the proposed read channel versus two other read channels.}
\vspace*{-0.5em}
\label{fig:BER} 
\end{figure}

The simulations are performed on a data set provided by data storage institute \cite{GFP_Kheong}. The waveforms are generated from the grain-flipping probability model in building the magnetized medium. This model generates realistic 2-D waveforms with media noise. A write frequency offset of $\tau_k^{(2)} = k\Delta T_2/T=2\times 10^{-4}k$, where $1/T$ is the ADC sampling rate, is injected into the bits of TRACK $2$ by linearly shifting the position of the writer. The rest of the tracks are written without any timing offsets. Fig. \ref{fig:BER} shows the BER performance of the ultimate proposed read channel in comparison with two other read channels. The proposed read channel is a cascade of the proposed neural network equalizer and the ROTAR detector \cite{ROTAR_JSAC}. The figure plots the average BER performance for the two middle tracks being detected using two readers separated based on track pitch (width) (TP). The proposed read channel is trained anew for each reader spacing to find the optimum target and equalizer pair for different readback waveforms selected.

The curve labeled ``GPRML, NN EQUALIZER" is the performance of the proposed read channel that shows a maximum of $37\%$ reduction in BER compared to the same read channel where a linear FIR equalizer replaces the neural network and is labeled as ``GPRML, LINEAR EQUALIZER". Further, the proposed read channel shows a maximum of $65\%$ reduction in BER compared to the conventional read channel, labeled as ``PRML, LINEAR EQUALIZER", that separately detects the two tracks using a pair of two MISO equalizers followed by two independent Viterbi detectors. In particular, to detect TRACK $1$ with $\Delta T_1=0$, a conventional GPR equalization strategy is used and is followed by a $2$-state Viterbi detector. To detect TRACK $2$ with $\Delta T_2\not=0$, a conventional GPR equalization is used and is followed by a conventional synchronization block that synchronizes to the timings of TRACK $2$. Finally, a $2$-state Viterbi detector detects the bits on TRACK $2$.

%


\vspace{-1em}

\bibliographystyle{IEEEtran}
\bibliography{IEEEabrv,refs}

\begin{thebibliography}{10}
\providecommand{\url}[1]{#1}
\csname url@samestyle\endcsname
\providecommand{\newblock}{\relax}
\providecommand{\bibinfo}[2]{#2}
\providecommand{\BIBentrySTDinterwordspacing}{\spaceskip=0pt\relax}
\providecommand{\BIBentryALTinterwordstretchfactor}{4}
\providecommand{\BIBentryALTinterwordspacing}{\spaceskip=\fontdimen2\font plus
\BIBentryALTinterwordstretchfactor\fontdimen3\font minus
  \fontdimen4\font\relax}
\providecommand{\BIBforeignlanguage}[2]{{%
\expandafter\ifx\csname l@#1\endcsname\relax
\typeout{** WARNING: IEEEtran.bst: No hyphenation pattern has been}%
\typeout{** loaded for the language `#1'. Using the pattern for}%
\typeout{** the default language instead.}%
\else
\language=\csname l@#1\endcsname
\fi
#2}}
\providecommand{\BIBdecl}{\relax}
\BIBdecl

\bibitem{EBS_thesis}
\BIBentryALTinterwordspacing
E.~{Banan Sadeghian}, ``{Synchronization and Detection for Two-Dimensional
  Magnetic Recording},'' Ph.D. dissertation, School of Elect. and Comp. Eng.,
  Georgia Inst. of Tech., Atlanta, GA, USA, 2016. [Online]. Available:
  \url{http://hdl.handle.net/1853/58210}
\BIBentrySTDinterwordspacing

\bibitem{APRTV_ICC}
E.~{Banan Sadeghian} and J.~R. Barry, ``{Partial-Response Maximum-Likelihood
  Joint Detection of Asynchronous Tracks},'' in \emph{IEEE International
  Conference on Communications (ICC 2021)}, 2021, pp. 1--6.

\bibitem{GPRML_TCOM}
{E. Banan Sadeghian} and J.~R. Barry, ``{Asynchronous Multitrack Detection With
  a Generalized Partial-Response Maximum-Likelihood Strategy},'' \emph{IEEE
  Transactions on Communications}, vol.~70, no.~3, pp. 1595--1605, 2022.

\bibitem{ROTAR_JSAC}
E.~Banan~Sadeghian and J.~R. Barry, ``{The Rotating-Target Algorithm for
  Jointly Detecting Asynchronous Tracks},'' \emph{IEEE Journal on Selected
  Areas in Communications}, vol.~34, no.~9, pp. 2463--2469, Sep. 2016.

\bibitem{Nair_nn1}
S.~K. {Nair} and {Jaekyun Moon}, ``{Simplified nonlinear equalizers},''
  \emph{IEEE Transactions on Magnetics}, vol.~31, no.~6, pp. 3051--3053, Nov.
  1995.

\bibitem{Nair_nn3}
S.~K. {Nair} and J.~{Moon}, ``{A theoretical study of linear and nonlinear
  equalization in nonlinear magnetic storage channels},'' \emph{IEEE
  Transactions on Neural Networks}, vol.~8, no.~5, pp. 1106--1118, Sep. 1997.

\bibitem{Ober_nn}
F.~{Obernosterer}, W.~F. {Oehme}, and A.~{Sutor}, ``Application of a neural
  network for detection at strong nonlinear intersymbol interference,''
  \emph{IEEE Transactions on Magnetics}, vol.~33, no.~5, pp. 2794--2796, Sep.
  1997.

\bibitem{Shen_CISS}
J.~{Shen} and N.~{Nangare}, ``{Nonlinear Equalization for TDMR Channels Using
  Neural Networks},'' in \emph{2020 54th Annual Conference on Information
  Sciences and Systems (CISS)}, Princeton, NJ, Mar. 2020, pp. 1--6.

\bibitem{Sayyafan_nn}
A.~{Sayyafan}, B.~J. {Belzer}, K.~{Sivakumar}, J.~{Shen}, K.~S. {Chan}, and
  A.~{James}, ``{Deep Neural Network Based Media Noise Predictors for Use in
  High-Density Magnetic Recording Turbo-Detectors},'' \emph{IEEE Transactions
  on Magnetics}, vol.~55, no.~12, pp. 1--6, Dec. 2019.

\bibitem{Siegel_RNN}
{Simeng Zheng}, {Yi Liu}, and {Paul H. Siegel}, ``{PR-NN: RNN-Based Detection
  for Coded Partial-Response Channels},'' \emph{IEEE Journal on Selected Areas
  in Communications}, vol.~39, no.~7, pp. 1967--1982, Jul. 2021.

\bibitem{GFP_Kheong}
K.~S. Chan, R.~Radhakrishnan, K.~Eason, M.~R. Elidrissi, J.~J. Miles,
  B.~Vasi\'{c}, and A.~R. Krishnan, ``{Channel Models and Detectors for
  Two-Dimensional Magnetic Recording},'' \emph{IEEE Transactions on Magnetics},
  vol.~46, no.~3, pp. 804--811, Mar. 2010.

\end{thebibliography}

\end{document}